\documentclass[reprint, 11pt,superscriptaddress,preprintnumbers,nofootinbib,aps]{revtex4}

\usepackage{graphicx}
\usepackage{amsmath,amssymb,amsfonts,amsthm,stmaryrd,mathtools,mathbbol,bm,physics,tensor}
\usepackage{color}
\usepackage{tikz}
\allowdisplaybreaks[1]

\usepackage[bookmarks,linktocpage, colorlinks=true, plainpages = false, citecolor = treegreen,  linkcolor=darkblue, urlcolor = darkblue, filecolor = blue]{hyperref} 

\def\be#1\ee{\begin{align}#1\end{align}}

\def\ba{\begin{eqnarray}}
	\def\ea{\end{eqnarray}}
\def\nn{\nonumber}

\definecolor{darkblue}{rgb}{0., 0.4, 0.8}
\definecolor{cadmiumred}{rgb}{1., 0., 0.22}
\definecolor{treegreen}{rgb}{0., 0.7, 0.3}
\definecolor{orchid}{rgb}{0.7., 0., 0.5}

\begin{document}

\title{Modified Friedmann equations and non-singular cosmologies in $d=4$ non-polynomial quasi-topological gravities}

\author{Johanna Borissova}
\email{j.borissova@imperial.ac.uk}
\affiliation{Abdus Salam Centre for Theoretical Physics, Imperial College London, London SW7 2AZ, UK}
\author{Jo\~ao Magueijo}
\email{j.magueijo@imperial.ac.uk}
\affiliation{Abdus Salam Centre for Theoretical Physics, Imperial College London, London SW7 2AZ, UK}

\begin{abstract}
	\bigskip
	{\sc Abstract:}
Quasi-topological theories of gravity are known to resolve black-hole singularities. We investigate whether the same mechanism can remove cosmological singularities. Focusing on non-polynomial curvature quasi-topological gravities in $d=4$ dimensions, we find three generic scenarios with the correct infrared limit but without a Big-Bang singularity, for universes filled with pure radiation or other standard matter. The first scenario yields a universe emerging from a de Sitter phase, a case for which the curvature invariants remain finite but the matter density diverges, albeit only at infinite affine distance.   The second one corresponds to a bouncing universe, and requires a multi-valued Lagrangian.
The third possibility is an asymptotically Minkowski origin, reminiscent of an eternally loitering universe. The matter energy density for this solution is non-singular even at infinite affine distance and does not enter a super-Planckian regime, but is instead approximately constant for the past eternity.

\end{abstract}

\maketitle
\tableofcontents

\section{Introduction}

Singularities are a generic prediction of classical general relativity. Under very general assumptions regarding the matter content and causal structure of spacetime, the singularity theorems imply the existence of geodesic incompleteness in both gravitational collapse and cosmological evolution~\cite{Penrose:1964wq,Hawking:1970zqf,Hawking:1973uf}. In particular, the standard Friedmann--Lemaître--Robertson--Walker (FLRW) solutions describing homogeneous and isotropic universes generically exhibit an initial Big-Bang singularity where curvature invariants and the matter density diverges. Resolving such singularities is therefore one of the central motivations for considering modifications of Einstein gravity in the ultraviolet (UV), while preserving its infrared (IR) successes.\\

A compelling framework for analysing the resolution of black-hole singularities through a purely gravitational mechanism are polynomial curvature quasi-topological gravities, polynomial higher-curvature extensions of general relativity, which exist in $d\geq 5$ dimensions and among other properties feature second-order equations on spherically symmetric backgrounds~\cite{Oliva:2010eb,Myers:2010jv,Dehghani:2011vu,Cisterna:2017umf,Bueno:2019ltp,Bueno:2019ycr, Bueno:2022res,Moreno:2023rfl,Bueno:2025qjk, Bueno:2024dgm,Frolov:2024hhe,Bueno:2024zsx,Aguayo:2025xfi,Bueno:2025qjk,Bueno:2025tli,Frolov:2026rcm,Tsuda:2026xjc,Li:2026mam}. In particular, \cite{Bueno:2024dgm} has derived static spherically symmetric regular black holes as gravitational vacuum solutions to such theories involving an infinite tower of curvature terms.~\footnote{For follow-up works and applications in the context of $d\geq 5$ polynomial curvature quasi-topological gravities, see e.g.~\cite{DiFilippo:2024mwm,Frolov:2024hhe,Bueno:2024zsx,Bueno:2025gjg,Aguayo:2025xfi,Bueno:2025qjk,Bueno:2025tli,Hao:2025utc,Frolov:2026rcm,Tsuda:2026xjc,Li:2026mam,Bueno:2026dln}.} By contrast, if the action is allowed to depend non-polynomially on curvature invariants, static spherically symmetric regular black holes can arise as vacuum solutions also to non-polynomial curvature quasi-topological theories in $d=4$ dimensions~\cite{Borissova:2026wmn,Borissova:2026krh,Bueno:2025zaj,Colleaux:2019ckh,Colleaux:2017ibe}. In fact, the solution space of these latter types of theories, which exist in any dimension $d\geq 4$, is notably larger than the one of polynomial curvature quasi-topological gravities~\cite{Borissova:2026wmn,Borissova:2026krh}.~\footnote{For first applications of this novel class of $d\geq 4$ non-polynomial curvature quasi-topological theories, see e.g.~\cite{Borissova:2026rbi,Konoplya:2026gim,Lutfuoglu:2026gis,Dubinsky:2026wcv}.} Such polynomial and non-polynomial quasi-topological theories are deeply interconnected with particular subclasses of integrable  two-dimensional Horndeski theories~\cite{Borissova:2026krh,Borissova:2026wmn, Borissova:2026dlz,Bueno:2025qjk,Bueno:2025gjg,Bueno:2025zaj,Colleaux:2019ckh}. From an opposite point of view, two-dimensional Horndeski theory provides a rich effective second-order geometrodynamic framework for $d\geq 4$ spherically symmetric spacetimes beyond general relativity~\cite{Carballo-Rubio:2025ntd}. See in particular~\cite{Boyanov:2025pes,Borissova:2026dlz} for applications to non-singular gravitational collapse processes.

 Remarkably, it has been shown that generic two-dimensional Horndeski theories can arise   
 from the spherical reduction of a purely gravitational theory in $d\geq 4$ dimensions~\cite{Borissova:2026krh}, see also the related discussions in~\cite{Colleaux:2017ibe,Colleaux:2019ckh}. Consequently, solving the equations of motion of two-dimensional Horndeski theory amounts to solving a corresponding higher-dimensional gravitational theory on spherically symmetric backgrounds. This is the key observation that motivates a detailed analysis of two-dimensional Horndeski theories as reduced higher-dimensional theories for the degrees of freedom of spherically symmetric spacetimes in various contexts also beyond black-hole physics.\\

In particular, these recent developments naturally raise the question of whether the same mechanism that resolves black-hole singularities in quasi-topological gravities~\cite{Bueno:2024dgm,Borissova:2026wmn,Borissova:2026krh}, thereby providing a promising route towards a non-singular paradigm for black-hole physics~\cite{Carballo-Rubio:2025fnc,Buoninfante:2024oxl}, can also eliminate cosmological singularities. 
Addressing the question of cosmological singularity resolution directly in higher-dimensional higher-curvature theories is often technically challenging, when not outright plagued by inconsistencies and ill-posedness in the presence of higher-derivative terms in the modified Friedmann equations. In this context, concerning the physically relevant case of $d=4$ spacetime dimensions, which we will primarily focus on here, let us mention that there exist polynomial higher-curvature theories beyond general relativity whose equations of motion are of second order in derivatives of the scale factor on FLRW cosmological backgrounds, such as the theories considered in~\cite{Arciniega:2018fxj,Cisterna:2018tgx,Arciniega:2018tnn,Moreno:2023arp}. The cosmological Einsteinian cubic curvature gravities considered in~\cite{Arciniega:2018fxj} admit solutions with a power-law geometric inflationary period, i.e., without the interplay of additional matter fields, connecting to a matter dominated era and followed by a period of late-time acceleration. Such instances of higher-curvature theories with second-order equations on FLRW backgrounds were extended notably  in~\cite{Arciniega:2018tnn} to infinite-curvature theories with this special property. These theories moreover have second-order linearised equations around generic maximally symmetric backgrounds, i.e., they do not propagate additional modes beyond the massless graviton,  and admit black-hole solutions satisfying $g_{tt}g_{rr}=-1$ in Schwarzschild gauge. It was shown in~\cite{Arciniega:2018tnn} that for the early-time geometric inflationary period to be governed by a de Sitter exponential growth rather than a power law, an infinite tower of curvatures in the action is required. This result aligns with later results in the context of static spherically symmetric black holes in polynomial quasi-topological gravities in $d\geq 5$ dimensions, whereby the limit of infinite-curvature quasi-topological densities is required for singularity resolution. In fact, the theory-dependent generating function entering the modified Friedmann equations derived from the $d=4$ cosmological gravities in~\cite{Arciniega:2018tnn} is a power series in one of the scalar invariants characterising the Riemann tensor (in a spatially flat universe, the squared Hubble parameter), just as in polynomial curvature quasi-topological gravities in $d\geq 5$, see~e.g.~\cite{Bueno:2025gjg,Sueto:2026epz}. In this sense, the above $d=4$ polynomial higher-curvature theories can be viewed as the naive dimensional reduction of $d\geq 5$ polynomial quasi-topological gravities within the subspace of FLRW cosmological backgrounds. A similar viewpoint forms the basis for the $d=4$ non-polynomial quasi-topological gravities considered in~\cite{Bueno:2025zaj}, which are constructed in such a way that their field equations on generic spherically symmetric backgrounds reproduce the $d=4$ analogue of the field equations in $d\geq5 $ polynomial quasi-topological gravities.~\footnote{This holds only up to some caveats, e.g.,~the reduced field equations derived from the $d=4$ non-polynomial quasi-topological gravities in~\cite{Bueno:2025zaj} receive no contribution from the effective order two in the curvature. This issue does not arise for the $d=4$ non-polynomial quasi-topological gravities constructed in~\cite{Borissova:2026wmn}.} Different from~\cite{Arciniega:2018fxj,Cisterna:2018tgx,Arciniega:2018tnn}, these theories possess second-order equations also on spherically symmetric backgrounds which are not of FLRW type.\\

In this work, we will  investigate the fate of the Big-Bang singularity in non-polynomial curvature quasi-topological gravities, focusing mainly on the physically relevant case of $d=4$ dimensions. Such theories are characterised by possessing second-order equations of motion on generic spherically symmetric backgrounds and admitting static spherically symmetric solutions satisfying $g_{tt}g_{rr}=-1$ in Schwarzschild gauge, see~\cite{Borissova:2026krh,Borissova:2026wmn} for a general classification of this landscape of theories. Instances of non-polynomial four-dimensional gravities with second-order equations on spherically symmetric backgrounds have been constructed, e.g., in~\cite{Deser:2007za,Colleaux:2017ibe,Colleaux:2019ckh,Bueno:2025zaj,Borissova:2026krh,Borissova:2026wmn}. Some first investigations of FLRW cosmological sectors for these types of theories have been performed in~\cite{Colleaux:2017ibe,Colleaux:2019ckh,Bueno:2025zaj,Ling:2025ncw}, see also~\cite{Bueno:2025gjg,Bueno:2026dln} for results concerning polynomial curvature quasi-topological gravities in higher dimensions which are relevant to our discussion, as the reduced theories belong to the same subclass of two-dimensional Horndeski theories. Moreover, let us also mention that the four-dimensional gravities constructed from the dimensional regularisation of Lovelock invariants on cosmological backgrounds feature an analogue equation of motion as the one we will be dealing with here~\cite{Fernandes:2025fnz}. In this work, we will provide a more complete geometric analysis of the different possibilities of Big-Bang singularity resolution in generic curvature quasi-topological gravities. These will turn out to be quite naturally dynamical realisations of different kinematic scenarios for cosmological singularity resolution discussed in~\cite{Carballo-Rubio:2024rlr}.\\

Roughly, the working mechanism can be described as follows. The spherical reduction of pure-curvature quasi-topological gravities yields particular subclasses of two-dimensional Horndeski theories characterised by two generating functions of one of the invariants determining the Riemann tensor for these spacetimes~\cite{Borissova:2026wmn,Borissova:2026krh}. We will assign a cosmological sector to these gravitational theories by evaluating the reduced theory on an FLRW background, see for instance the discussion in~\cite{Colleaux:2017ibe,Colleaux:2019ckh}. In our case, the resulting equations of motion will be characterised entirely by a single one-variable generating function, as for the four-dimensional non-polynomial quasi-topological gravities constructed in~\cite{Bueno:2025zaj}. Focusing on universes with spatial curvature $k=0$, the cosmological equations reduce to a single algebraic relation between the energy density $\rho$ of the external matter and a function $h\qty(H^2)$ of the squared Hubble parameter. This equation plays the role of a generalised Friedmann equation and provides a convenient starting point for exploring non-singular cosmological evolutions.

We will show that requiring the correct IR limit compatible with general relativity, while avoiding a Big-Bang singularity, leads to three qualitatively distinct possibilities. A first possibility is that the universe emerges from an asymptotic de Sitter phase even when the matter content is pure radiation. In this case curvature invariants remain finite but the matter density diverges at infinite affine distance. The second one corresponds to a bouncing cosmology, in which the scale factor attains a non-zero minimum. This type of effective geometry arises notably in loop quantum cosmology~\cite{Bojowald:2001xe,Ashtekar:2006rx,Ashtekar:2007em,Singh:2014fsy}. We demonstrate that realising such a solution geometrically requires the underlying Lagrangian to be multi-valued, a feature that has not been previously emphasised. Finally, a third scenario arises in which the universe originates from an asymptotically Minkowski state. This solution is completely non-singular and avoids the super-Planckian regime encountered in the de Sitter case. Such a scenario has been considered for instance in the context of asymptotic safety in~\cite{Bonanno:2017gji} and more generally in studies of non-singular cosmologies performed in~\cite{Lesnefsky:2022fen,Easson:2024uxe,Easson:2024fzn}. The novelty here is that we are here considering these non-singular cosmologies as derived from a Lagrangian principle, the existence of which is a priori not guaranteed at the level of the quantum-corrected effective field equations advocated in any of the above quantum-gravity approaches.

Remarkably, the idea of modifying Einstein gravity to include non-polynomial higher-curvature terms dates back to original proposals of cosmological singularity resolution in the context of Markov's limiting curvature hypothesis~\cite{Markov:1982rcm}, according to which curvature invariants for every solution must be universally bounded by inverse powers of a fundamental length scale.~\footnote{Another early discussion of non-polynomial gravities has been  motivated by Born-Infeld electrodynamics in~\cite{Deser:1998rj}.} In the limit in which this bound is saturated, the universe approaches a de Sitter state. In~\cite{Mukhanov:1991zn,Brandenberger:1993ef,Brandenberger:1995hd}, Brandenberger, Mukhanov et al proposed an implementation of the limiting curvature hypothesis at the level of the effective gravitational action by means of Lagrange multiplier fields ensuring the boundedness of a given set of curvature invariants together with the condition that the metric approaches the de Sitter metric upon saturation of this bound. Requiring simultaneously an infrared limit compatible with general relativity naturally leads to non-polynomial potentials for the auxiliary fields, whose derivatives determine the functions of curvature invariants entering the gravitational action which are necessary for a realisation of the limiting curvature condition onshell. This implementation of the limiting curvature condition leads to a singularity-free expanding universe emerging from a de Sitter phase, whereby the period of inflation is purely gravitationally driven. In particular, the matter couplings in these theories go to zero as the curvature approaches its limiting value. A similar purely gravitational singularity resolution mechanism lies also at the core of our analysis, i.e., instead of achieving singularity resolution by a violation of energy conditions in the matter sector, the gravitational dynamics at early times gets modified.

In relation to this, let us emphasise that, in this work, by considering two-dimensional Horndeski theories as reduced higher-dimensional gravities~\cite{Borissova:2026krh}, we are targeting specifically the modifications to the gravitational dynamics stemming from (in general non-polynomial) higher-curvature and higher-derivative terms in the action, whereas we will keep the matter sector of the field equations to be a simple perfect fluid. One may also consider the question of singularity resolution in the presence of a more extensively selected matter sector, such as for instance an electromagnatic field involving higher derivatives and non-linearities. Given that non-linear electromagnetic fields are well-known to source some regular black holes when coupled to general relativity, e.g.~\cite{Ayon-Beato:1998hmi,Ayon-Beato:1999qin,Ayon-Beato:1999kuh,Ayon-Beato:2000mjt,Bronnikov:2000vy,Dymnikova:2004zc,Bronnikov:2022ofk,Balart:2014cga,Fan:2016hvf,Rodrigues:2018bdc}, it would be interesting to understand to which extent black-hole and cosmological singularity resolution in gravitational theories with second-order equations on spherically symmetric backgrounds, in particular in quasi-topological gravities, is a combined effect of gravity and matter. In this context, let us mention first derivations of charged black holes in $d\geq 5$ polynomial curvature quasi-topological gravities coupled to Born-Infeld-type non-linear electrodynamics, which describe fully regular gravitational and electromagnetic fields~\cite{Hennigar:2017ego,PinedoSoto:2026hfm}. At the same time, regular geometries can be sourced in polynomial quasi-topological gravities also from singular minimally coupled matter, or can be regularised by the inclusion of non-minimal couplings~\cite{Bueno:2026dln}. The status of coupling these gravitational theories to matter, in what concerns black-hole and cosmological singularity resolution, is thus at this stage rather unsettled. It is, however, a conceptually important question to address from the viewpoint of a consistent gravity-matter effective theory.
\\

The remainder of this paper is organised as follows. In Sec.~\ref{Sec:2DHorndeski} we introduce two-dimensional Horndeski theories as reduced gravitational theories, and focus on the subclass of two-dimensional Horndeski theories which can arise from the reduction of curvature quasi-topological gravities. We evaluate the equations of motion for an FLRW cosmological ansatz. In particular, this section states the master equation which serves as our starting point for the reconstruction of non-singular cosmologies. 
Then, in Sec.~\ref{Sec:Acceleration} we derive the conditions on $h\qty(H^2)$ under which a period of accelerated expansion can arise without violating the strong energy condition. This is because such acceleration is kinematically required in the three distinct scenarios for singularity resolution produced by the modified gravitational dynamics. In Sec.~\ref{Sec:NonSingularCosmologies},
 we proceed to investiate three separate scenarios for Big-Bang singularity resolution by analysing corresponding exemplary choices for the funtion $h$ --- an early-time de Sitter phase in Subsec.~\ref{SecSub:deSitter}, a bouncing cosmology in Subsec.~\ref{SecSub:Bounce}, and an early-time Minkowski phase in Subsec.~\ref{SecSub:Minkowski}. We finish with a conclusion in Sec.~\ref{Sec:Conclusions}.

\section{2D Horndeski theory for FLRW cosmologies}\label{Sec:2DHorndeski}

In the following we will consider two-dimensional Horndeski theory as a reduced gravitational theory for the degrees of freedom of $d$-dimensional warped-product spacetimes
\ba\label{eq:MetricWarped}
g_{\mu\nu}(x) \dd{x}^\mu \dd{x}^\nu &=& q_{ab}(y)\dd{y}^a \dd{y}^b + \varphi(y)^2 \dd{\Sigma_{d-2}^2}\,,
\ea
where $\dd{\Sigma_{d-2}^2} $ is the surface element of a $d-2$ dimensional compact space of constant sectional curvature. In fact, it has been shown that generic two-dimensional Horndeski theories can arise as the reduction on~\eqref{eq:MetricWarped} of purely  gravitational generally covariant  theories in $d\geq 4 $ dimensions~\cite{Borissova:2026krh}, see also~\cite{Colleaux:2019ckh}, and therefore solving the equations of motion of two-dimensional Horndeski theory amounts to actually solving such a gravitational theory. \\

We will be interested specifically in FLRW cosmologies with line element
\ba\label{eq:Metric}
\dd{s}^2 &=& - \dd{\tau }^2 + a(\tau)^2\qty[\frac{\dd{r}^2}{1-k r^2}+  r^2 \dd{\Omega^2_{d-2}}]\,,
\ea
corresponding to spherical reduction, and the identification
\ba
q_{ab}(y)\dd{y}^a \dd{y}^b  &=&  - \dd{\tau }^2 + a(\tau)^2\frac{\dd{r}^2}{1-k r^2}\,,\label{eq:q} \\
\varphi(y) &=& a(\tau) r \label{eq:Varphi}\,.
\ea
The general two-dimensional Horndeski action for a metric $q_{ab}(y)$ and a scalar field $\varphi(y)$ can be written as~\cite{Horndeski:1974wa,Kobayashi:2011nu,Kobayashi:2019hrl}
\ba\label{eq:SHorndeskiLong}
S_{\text{Horndeski}}[q,\varphi] &=& \int \dd[2]{y} \sqrt{-q}\,\Big[h_2(\varphi,\chi) - h_3(\varphi,\chi) \Box \varphi + h_4(\varphi,\chi)\mathcal{R} \nn\\
&-& 2 \partial_\chi h_4(\varphi,\chi) \qty[\qty(\Box \varphi)^2 - \nabla_a \nabla_b \varphi \nabla^a \nabla^b \varphi]\Big] \,,\quad \quad 
\ea
where $\mathcal{R}$ is the Ricci scalar and $\chi = \nabla_a \varphi \nabla^a \varphi$ is the scalar field kinetic term. For the above choice of ansatz, the latter is given by
\ba\label{eq:Chi}
\chi &=& \nabla_a \varphi \nabla^a \varphi \,\, = \,\, 1 - \qty[k^2 +\dot{a}^2]r^2 \,.
\ea
The equations of motion for the metric $q_{ab}(y)$ derived from~\eqref{eq:SHorndeskiLong} in the presence of a two-dimensional external matter source $T_{ab}$ can be written as~\cite{Carballo-Rubio:2025ntd}
\ba
\mathcal{E}_{ab} &=& - \frac{1}{2}\big[\alpha + 2\beta \Box \varphi\big]q_{ab}  +
\omega \nabla_a \varphi \nabla_b \varphi  +  \beta \nabla_a \nabla_b \varphi \,\, = \,\, T_{ab}\,, \label{eq:Eab}
\ea
where
\ba
\alpha(\varphi,\chi) \,\,= \,\,  h_2 + \chi \partial_\varphi \qty( h_3  - 2 \partial_\varphi h_4)\,, \,\,\, \quad \quad 
\beta(\varphi,\chi) \,\,=\,\, \chi \partial_\chi \qty(h_3  - 2 \partial_\varphi h_4) - \partial_\varphi h_4 \,,
\label{eq:AlphaBeta}
\ea
and
\ba\label{eq:omega}
\omega(\varphi,\chi) &=& \partial_\chi \alpha - \partial_\varphi \beta \,.
\ea
Evaluating the independent components of the equations of motion tensor~\eqref{eq:Eab} for the ansatz~\eqref{eq:Metric}, results in
\ba
\mathcal{E}_{tt} &=& \frac{1}{2}\alpha - \frac{k+\dot{a}^2}{a} r \beta + r^2 \dot{a}^2 \omega \,\, =\,\, T_{tt}\,,\\
\mathcal{E}_{tr} &=& r a \dot{a}  \omega \,\, = \,\, T_{tr}\,,\\
\mathcal{E}_{rr} &=&-\frac{1}{2} \frac{a^2}{1-kr^2} \qty[ \alpha - 2 r \ddot{a} \beta ] + a^2 \omega\,\,=\,\, T_{rr}\,,
\ea
where we have omitted the explicit arguments of the functions $\alpha(\varphi,\chi)$ and $\beta(\varphi,\chi)$, which onshell are given by~\eqref{eq:Varphi} and~\eqref{eq:Chi}. We will assume that the effective two-dimensional source arises as the reduction on~\eqref{eq:Metric} of a $d$-dimensional energy-momentum tensor for a perfect fluid with energy density $\rho$ and pressure $p$,
\ba
\mathcal{T}^{\mu\nu} &=& (\rho +p) u^\mu u^\nu + p g^{\mu\nu} \,,
\ea 
obeying the conservation equation 
\ba\label{eq:Conservation}
\dot{\rho} + (d-1)(\rho +p) \frac{\dot{a}}{a} &=& 0\,.
\ea
In this case, the offshell generalised Bianchi identity implied by general covariance of the reduced action ensures that onshell of the equations of motion for the metric $q_{ab}(y)$, the equation of motion for the scalar field $\varphi(y)$ is automatically satisfied. Concretely, adopting the standard Hilbert prescription for the definition of the energy-momentum tensor $T_{ab}$ from a reduced matter action $S_{m}$, the variations $\mathcal{E}_{ab}$ and $\mathcal{E}_{\varphi}$ of the Horndeski action with respect to $q^{ab}$ and $\varphi$ satisfy
\ba
\nabla^a \mathcal{E}_{ab} + \frac{1}{2}\mathcal{E}_{\varphi} \nabla_b \varphi &=& \nabla^a T_{ab}\,,
\ea
showing that in this case solving $\mathcal{E}_{ab}=T_{ab}$ is sufficient to guarantee $\mathcal{E}_{\varphi}=0$.\\

The effective two-dimensional source obtained from the reduction of the $d$-dimensional one is determined by~\cite{Carballo-Rubio:2025ntd}
\ba
T_{ab} &=& \varphi^{d-2} \mathcal{T}_{\mu\nu} \delta_a^\mu \delta_b^\nu\,,
\ea
i.e, such that in particular $T_{tr}=0$. Therefore, there is no external source on the right-hand side of the $tr$-equation equation of motion. It follows that, as long as $\dot{a}$ is non-zero, it must be $\omega =0$. In other words, for cosmological configurations~\eqref{eq:Metric}, i.e, for~\eqref{eq:q} and~\eqref{eq:Varphi}, to be a solution to a two-dimensional Horndeski theory, supplemented by our particular choice of external matter content, this theory must satisfy the integrability condition
\ba\label{eq:Integrability}
\omega(\varphi,\chi) &=& \partial_\chi \alpha - \partial_\varphi \beta  \,\, = \,\, 0\,.
\ea
This condition defines the most general notion of  quasi-topological gravities~\cite{Borissova:2026krh}.\\

In this work, we will focus on quasi-topological gravities built non-polynomially from curvature invariants without covariant derivatives in $d=4$, such as the theories considered in~\cite{Borissova:2026wmn,Borissova:2026krh,Bueno:2025zaj}. Polynomial curvature quasi-topological gravities do not exist in $d=4$~\cite{Bueno:2019ltp,Moreno:2023rfl}. \\

Under the assumption of a pure-curvature Lagrangian density, the functions $\alpha$ and $\beta$ depend on the following eigenvalue of the Riemann tensor for the line element~\eqref{eq:Metric},
\ba
\psi &\equiv & \frac{1-\chi}{\varphi^2} \,\,=\,\,\frac{k + \dot{a}^2}{a^2} \,,
\ea
in a particular way established in~\cite{Borissova:2026krh,Borissova:2026wmn}. This scalar quantity is the central variable in terms of which the equations of motion for generic curvature quasi-topological gravities are written.  Notice that by setting $\omega = 0$, the $tt$- and $rr$-equations can be written as
\ba
\frac{1}{2}\alpha - \frac{k+ \dot{a}^2}{a^2} r a \beta \,\,= \,\, T_{tt}\,,\\
-\frac{1}{2} \frac{a^2}{1-k r^2} \qty[\alpha - 2 r \ddot{a}\beta]\,\,=\,\, T_{rr}\,.
\ea
The second equation can be replaced by the conservation equation~\eqref{eq:Conservation}.
Now let us consider the implications of the integrability condition~\eqref{eq:Integrability} as discussed in~\cite{Boyanov:2025pes,Carballo-Rubio:2025ntd,Borissova:2026dlz,Borissova:2026wmn,Borissova:2026krh}. This condition can be solved by means of a characteristic function $\Omega(\varphi,\chi)$ satisfying
\ba
\alpha(\varphi,\chi) \,\, = \,\, \partial_\varphi \Omega(\varphi,\chi)\,,\,\,\, \quad \quad \beta(\varphi,\chi) \,\,=\,\, \partial_\chi \Omega(\varphi,\chi)\,.
\ea
For this function, with $\varphi = a r$ and $\chi = 1 - \qty[k + \dot{a}^2]r^2$ onshell, it holds according to~\eqref{eq:Chi} that
\ba
\derivative{}{\varphi}\qty[\Omega(\varphi,\chi)] &=& \alpha(\varphi,\chi) + \beta(\varphi,\chi) \partial_\varphi \chi\,,
\ea
whereby, holding $\tau$ fixed,
\ba
\partial_\varphi \chi &=& \partial_r \chi \partial_\varphi r \,\,=\,\, - \frac{2 r}{a^2} \qty(k + \dot{a}^2)\,.
\ea
In other words, we may write the left-hand side of the $tt$-equation as a total derivative,
\ba
\frac{1}{2}\eval{\derivative{}{\varphi}\qty[\Omega(\varphi,\chi)] }_{\varphi = ar,\, \chi = 1- \qty[k + \dot{a}^2]r^2 } &=& T_{tt} \,.
\ea
It the absence of an external matter source on the right-hand side, this equation states that $\Omega$ must be equal to an integration constant onshell, which in the context of static spherically symmetric and asymptotically flat solutions to the $d$-dimensional theory, is related to the ADM mass~\cite{Borissova:2026krh,Borissova:2026wmn}. For non-zero $T_{tt}$, we can still integrate the above equation with respect to $\varphi$ to obtain
\ba\label{eq:OmegaRho}
\frac{1}{2}\Omega(\varphi,\chi) &=& \frac{1}{d-1} \varphi^{d-1} \rho\,.
\ea
We are now in a position to take into account the constraints on the functional form of the generating function $\Omega(\varphi,\chi)$ for two-dimensional Horndeski theories which can arise from the reduction of a curvature quasi-topological gravity in $d\geq 4$ dimensions~\cite{Borissova:2026krh,Borissova:2026wmn}. For generic $d$-dimensional non-polynomial theories of this type, the function $\Omega$, written in terms of the variables $\varphi$ and $\psi$, must be of the form
\ba\label{eq:Master}
\Omega(\varphi,\psi) &=& \qty[h(\psi) - \frac{d-2}{d-1}\psi g(\psi)] \varphi^{d-1} + g(\psi) \varphi^{d-3}\,,
\ea
for certain functions $h$ and $g$ defined in terms of the functions $h_i$ rescaled by $d$-dependent powers of $\varphi$~\cite{Borissova:2026krh}. Such theories exist and can be constructed explicitly in any dimension $d\geq 4$~\cite{Borissova:2026krh,Borissova:2026wmn}. In the particular case when this curvature quasi-topological gravity is built polynomially from curvature invariants, cf.~e.g.~\cite{Oliva:2010eb,Myers:2010ru,Myers:2010jv,Dehghani:2011vu,Cisterna:2017umf,Bueno:2019ltp,Bueno:2019ycr, Bueno:2022res,Moreno:2023rfl,Bueno:2024dgm,Frolov:2024hhe,Bueno:2024zsx,Aguayo:2025xfi,Bueno:2025qjk,Bueno:2025tli,Frolov:2026rcm,Tsuda:2026xjc,Li:2026mam}, the function $g(\psi)$ is zero and moreover the function $h(\psi)$ arises from the resummation of polynomial curvature quasi-topological densities evaluated on~\eqref{eq:Metric}, in the form~\cite{Bueno:2024dgm}
\ba\label{eq:hPol}
h(\psi) &=& \psi + \sum_{n=2}^N \alpha_n \psi^{n}\,,
\ea
where $\alpha_n$ are arbitrary coupling constants of dimensions $\qty[\text{length}]^{2(n-1)}$ in Planck units. 
The limit $N \to \infty$ amounts to the inclusion of an infinite tower of curvature terms in the action. These latter types of theories beyond general relativity exist only in $d \geq 5$ dimensions~\cite{Bueno:2019ltp,Moreno:2023rfl}, however, in~\cite{Bueno:2025zaj} four-dimensional non-polynomial generally covariant curvature have been constructed whose reduction produces polynomial second-order equations corresponding to a subset of the two-dimensional Horndeski theories which can arise from polynomial curvature quasi-topological densities in $d\geq 5$. For a more general discussion regarding the possibility of constructing such non-polynomial theories, as well as for an extension to generating generic two-dimensional Horndeski theories from non-polynomial gravities, see~\cite{Borissova:2026wmn,Borissova:2026krh,Colleaux:2019ckh,Colleaux:2017ibe}.\\

 Comparing~\eqref{eq:Master} with~\eqref{eq:OmegaRho}, we see that $g(\psi)$ must be set to zero  in the present setup. We are thus left to consider the equation
\ba\label{eq:EOM}
h(\psi) &=& \frac{2}{d-1} \rho\,,
\ea
whereby $h(\psi)$ for polynomial curvature quasi-topological gravities in $d\geq 5$ must arise in the form~\eqref{eq:hPol}, cf.~e.g.~\cite{Bueno:2024dgm}, but more generally for the constructions of non-polynomial curvature quasi-topological gravities in $d\geq 4$ considered in~\cite{Borissova:2026krh,Borissova:2026wmn}, the function $h(\psi)$  can be an arbitrary function which does not need to arise from such a resummation. Accordingly, in the following we will consider $h(\psi)$ to be a generic function.\\

In summary, reinstating factors of $8\pi G$ temporarily, the central equations for our further analyses  are the equation of motion~\eqref{eq:EOM} and the conservation equation for the stress-energy tensor of the fluid~\eqref{eq:Conservation}, i.e.,
\ba
h(\psi)&=& \frac{16 \pi G}{(d-2)(d-1)}\rho \,\, \equiv \,\, \varrho\,,\label{eq:EqMaster}\\
\dot{\rho} &=& - (d-1)\qty(\rho +p)\frac{\dot{a}}{a}\label{eq:EqConservation}\,,
\ea
where  $\varrho$ defined above has dimension $[\text{length}]^{-2}$ in Planck units. \\

Notice that in the particular case when $k=0$, the variable $\psi$ becomes the squared Hubble parameter,
\ba\label{eq:Psik0}
\psi &=& \qty(\frac{\dot{a}}{a})^2 \,\,\equiv \,\,H^2 \,.
\ea
Our goal will be to distinguish different geometric possibilities for the resolution of the Big-Bang singularity by their properties of the function $h$, whereby in this work we will restrict ourselves to $k=0$.
Making use of the equation of state $p=w\rho $ with constant parameter $w$,  the conservation equation~\eqref{eq:EqConservation} can be integrated to obtain the rescaled energy density $\varrho $ as a function of the scale parameter as follows,
\ba\label{eq:Rhoa}
\varrho (a) &=& \varrho_0 \qty(\frac{a}{a_0})^{(1-d)(1+w)}\,.
\ea
This will allow us, given an invertible function $h\qty(H^2)$, to solve~\eqref{eq:EqMaster} in the form $H^2(\varrho)$, and therefrom, by inserting~$\varrho(a)$, to compute the scale factor $a$ as a function of time $\tau$. In the following we will denote by $\varrho_0$ the value of $\varrho$ when $H^2 = H_0^2$.

\section{Conditions for acceleration}\label{Sec:Acceleration}

In everything that follows we  will always require a late-time infrared (IR) limit compatible with general relativity, i.e.,
\ba
\text{IR}: \quad \quad h\qty(H^2) \,\,\,\to \,\,\, H^2 \,,
\ea
in which case the solution to~\eqref{eq:EqMaster} for the scale factor is given by
\ba
\frac{\dd{a}}{a \sqrt{\varrho(a)}} &=& \dd{\tau} \,\,\, \quad \,\,\, \Rightarrow  \,\,\, \quad \,\,\, \frac{2}{(d-1)(1+w)} \,\qty(\frac{a}{a_0})^{\frac{1}{2}(d-1)(1+w)}\,\,=\,\, H_0 \qty(\tau - \tau_0)\,.
\ea
We will from now on mainly specialise to $d=4$ and an equation of state parameter $w=1/3$ appropriate for a radiation-dominated universe. In this case, at late times $\tau \to \infty$ we impose $a \propto \tau^{1/2}$.\\

Our goal is to analyse scenarios for the resolution of the cosmological singularity by means of modified gravity in the UV,  as encoded in the characteristic function $h$. Concretely, we will consider scenarios leading to an early de Sitter phase, a bounce, and a primordial static universe eternal in the past. Thinking purely in kinematic terms, i.e., designing functions $a(\tau)$ linking such behavior with the IR solution, this tells us that such scenarios must involve a period of accelerated expansion, either as the mechanism responsible for singularity resolution, or as a by-product. We therefore start by deriving conditions on the characteristic function $h$ for solutions to admit a phase of accelerated expansion, i.e., $\ddot{a}>0$.\\

To that end, by taking the time derivative of the equation of motion
\ba
h\qty(H^2) &=& \varrho \,,
\ea
using the conservation equation~\eqref{eq:EqConservation} and reusing the equation of motion, we find
\ba
    2H\dot H h' &=&
    -(d-1)(1+w)H h\,,
\ea
or equivalently, assuming $h' \neq 0$,
\ba
    \dot H=-\frac{(d-1)}{2} (1+w)\frac{h}{h'} \,.
\ea
Notice moreover the following identity, obtained by taking the time derivative of the Hubble parameter, 
\ba\label{ddota}
\frac{\ddot{a}}{a}&=&\dot H+H^2\,.
\ea
Therefrom we obtain
\ba\label{ddotamain}
    \frac{\ddot{a}}{a} &=& 
  H^2 \left(1-\frac{1}{2}(d-1)(1+w)\frac{h\qty(H^2)}{h'\qty(H^2) H^2} \right)\,,
\ea
setting out the condition for accelerated expansion, i.e., $\ddot{a}/a > 0$,
\ba
    1-\frac{1}{2}(d-1)(1+w)\frac{h\qty(H^2)}{h'\qty(H^2)H^2} &>&0\,,
\ea
which holds for a general perfect fluid with constant equation of state parameter $w$. \\

In the IR, general relativity amounts to $h\qty(H^2) = H^2$ and $h'\qty(H^2)=1$, and so the above condition for acceleration reduces to 
\ba 
w &<& -\frac{d-3}{d-1}\,.
\ea
This is just the usual requirement of violation of the strong energy condition in $d$ dimensions, which corresponds concretely to $\rho+3p<0$ for $d=4$. 

In the UV, the above condition for acceleration is controlled by the central quantity in the theory, $h/\qty(h'H^2)$, and becomes,
\ba\label{eq:acceleration}
\frac{h\qty(H^2)}{h'\qty(H^2) H^2} &<& \frac{2}{(d-1)(1+w)}\,.
\ea
E.g.~for $d=4$ and $w = 1/3$, as we will investigate later, the right-hand side equals $1/2$.\\

Hence, we can design the characteristic function $h\qty(H^2)$ so as to generate a phase of acceleration without breaking the strong energy condition, a requirement we shall impose in what follows. \\

What could $h$ do in the UV to lead to acceleration, discounting baroque behaviour?  First, consider the possibility that $h$ as a function of $H^2$ continually falls below the bisector $h\qty(H^2)= H^2$. In this case
$h/\qty(h'H^2)$ increases above one and this does not lead to acceleration. But suppose that $h$ continues to bend down until it reaches a maximum and subsequently decreases, thereby opening the door to a qualitatively different regime. However, in this case~\eqref{ddotamain} implies that if $h'$ vanishes at an isolated point, then $\ddot a$ diverges, signalling a ``sudden singularity''~\cite{Barrow:2010wh}. Since our aim is precisely to remove cosmological singularities, we exclude such pathological scenarios in this work, but  will treat these more leniently in future work.

Closing the door here to characteristic functions with such properties, and excluding contrived constructions in which $h'$ oscillates, we are therefore led to consider models in which $h$ eventually rises above the bisector $h\qty(H^2)=H^2$. The subsequent behaviour of $h$ then determines the three scenarios for singularity resolution that we will consider.

\section{Non-singular FLRW cosmologies}\label{Sec:NonSingularCosmologies}

\begin{figure}[t!]
	\centering
	\includegraphics[width=1\textwidth]{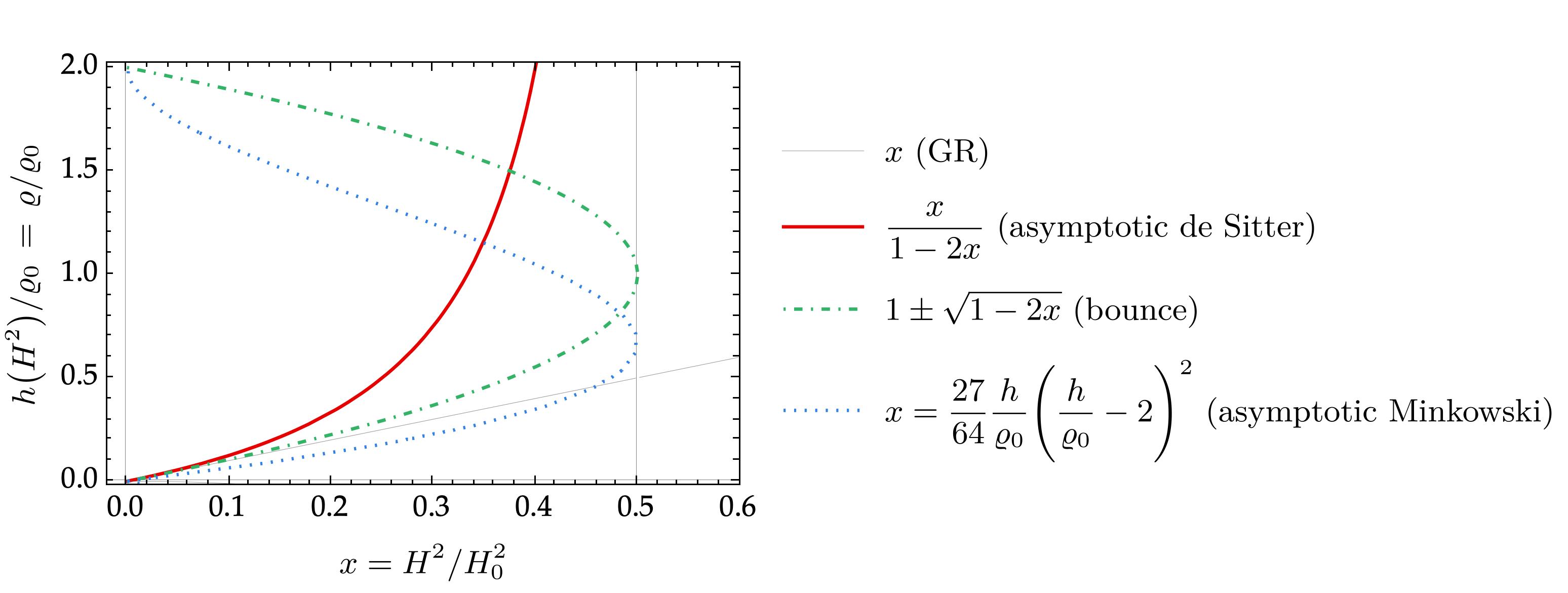}
	\caption{\label{Fig:PhaseDiagram} Examples of choices of characteristic functions $h\qty(H^2)$ resulting in Big-Bang singularity resolution via an asymptotic de Sitter phase, a bounce, or an asymptotic Minkowski phase.
	}
\end{figure}

In this section, as already announced, we will restrict attention to a radiation-dominated universe in the physically most relevant case of $d=4$ dimensions. In this case the condition for acceleration stated in~\eqref{eq:acceleration} becomes explicitly
\ba
    \frac{h\qty(H^2)}{h'\qty(H^2) H^2} &<& \frac{1}{2}\,,
\ea
and starts as soon as the inequality is saturated as a result of $h$ rising above $h\qty(H^2)=H^2$, decreasing the value of $h/\qty(h'H^2)$. If
$h$ continues to rise and reaches a constant finite $\gamma=h/\qty(h' H^2)$~\footnote{For any power law $a \propto \tau^n$ and $d=4$ as well as $w=1/3$, equation~\eqref{ddotamain} then becomes $n(n-1)=n^2(1-2\gamma)$, i.e., $n=1/2\gamma$.}, this leads to power-law acceleration,
\ba
    a &\propto & \tau^\frac{1}{2\gamma}\,,
\ea
which still has a Big-Bang singularity in the past. This is not surprising, since the curvature scalar $\psi$ in this case would not be capped but would increase linearly with $\varrho$, resulting in a  divergence of both  at $\tau=0$. 

We are therefore led to the first possibility for singularity resolution --- that $h$ be singular at some point, leading to a de Sitter past as considered kinematically in various contexts before, cf., e.g.,~\cite{Mukhanov:1991zn,Carballo-Rubio:2024rlr} and discussed in the context of four-dimensional non-polynomial curvature quasi-topological gravities in~\cite{Bueno:2025zaj}.
This makes sense, indeed, by
considering that the $h\qty(H^2)$ diagram is also a $\varrho\qty(H^2)$ diagram and that $\varrho\propto 1/a^4$, and so must diverge as $\tau\rightarrow -\infty$ in a universe with constant $H^2$. Thus, we conclude that for this scenario to be realised, $h\qty(H^2)$ must have a singularity at $H^2=H_0^2$, where $H_0$ characterizes the de Sitter asymptotic state (with $k=0$). 
We have depicted this scenario by the red curve in Fig.~\ref{Fig:PhaseDiagram}, which rises to infinity at $H^2/H_0^2=1/2$.

This leaves only one route to realise other scenarios, which is that $h$ be double-branched for $0<H^2<H_0^2$ and limited to this interval, for some $H_0$. A bounce (i.e., $H^2=0$ at an isolated point) means that the $h$-axis is hit again by this second branch, signaling $H^2=0$ for some $h=\varrho_0$. This is a necessary but not sufficient condition for a bounce, and we will check it with a concrete choice of $h$. We have depicted this scenario by the green curve in Fig.~\ref{Fig:PhaseDiagram}, which arises by combining the two branches of a specific choice of function $h$ leading to a bounce. 

The third scenario arises as a special case of the second, namely when in the UV branch the function hits the $h$-axis (i.e., $H^2=0$) with $h'(0)=\infty$. As~\eqref{ddotamain} shows, this implies $\ddot a=0$ when this point is reached, a necessary but not sufficient condition for an asymptotic Minkowski beginning. We have depicted an example of this scenario by the blue curve in Fig.~\ref{Fig:PhaseDiagram}.\\

The above considerations are merely indicative (i.e., they use necessary but not sufficient conditions), and specifically they do not tell us how the trajectories in $h\qty(H^2)$ are traversed in terms of proper (affine) time, an essential matter to evaluate singularity resolution and geodesic completeness. 
We will therefore examine concrete solutions corresponding to these three geometrically distinct cases.

\subsection{de Sitter asymptotic universe}\label{SecSub:deSitter}

\begin{figure}[t!]
	\centering
	\includegraphics[width=0.5\textwidth]{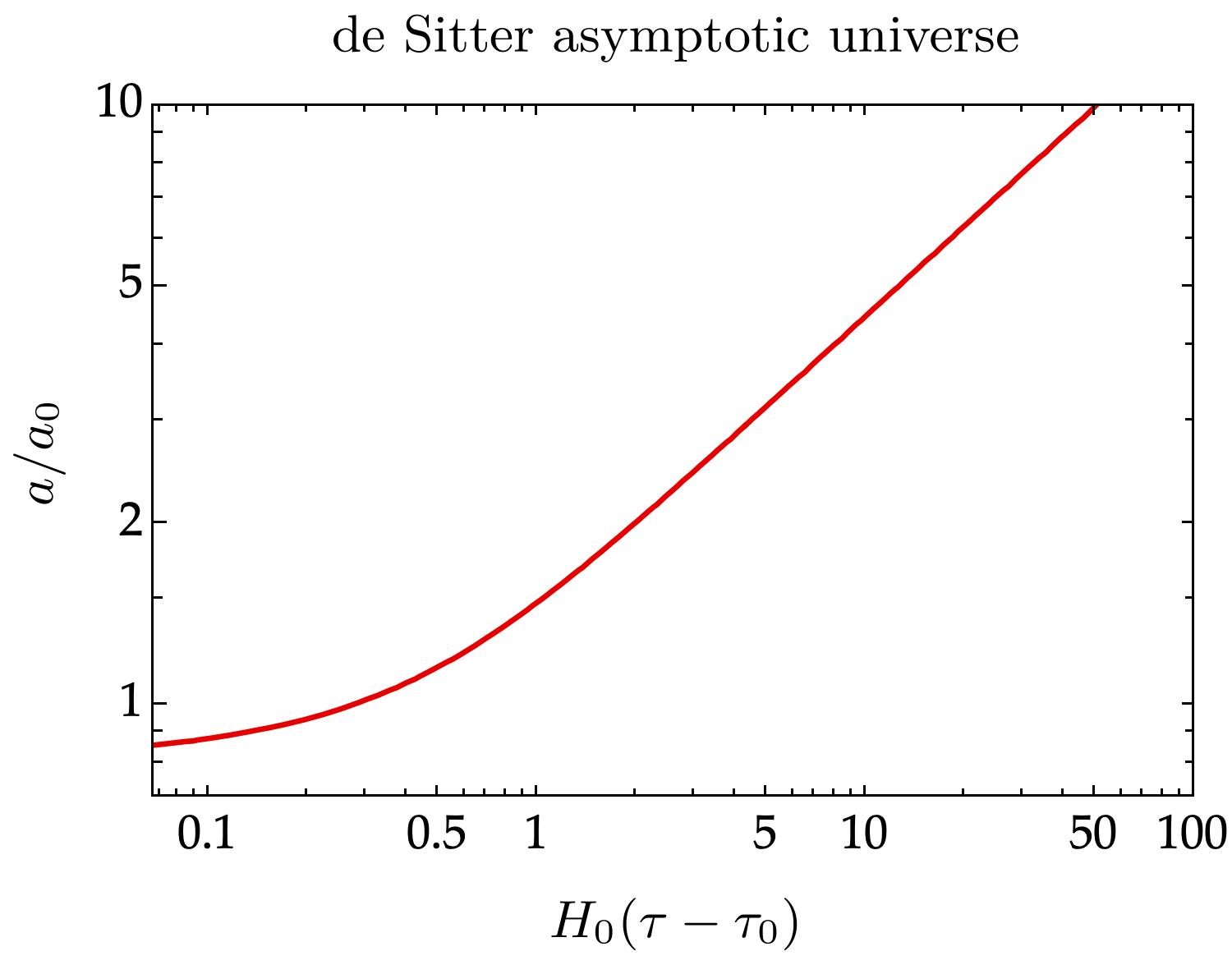}
	\caption{\label{Fig:deSitter} Scale factor as a function of time for a $d=4$ radiation-dominated FLRW universe with characteristic function $h$ in~\eqref{eq:hdeSitter} according to~\eqref{eq:adeSitter}. The curve shows a transition from exponential expansion at early times to a power-law scaling at late times.}
	
\end{figure} 
Following the above discussion, let us consider the characteristic function
\ba\label{eq:hdeSitter}
h\qty(H^2) &=& \frac{H^2}{1- \frac{H^2}{H_0^2}}\,,
\ea
which in the context of static spherically symmetric black holes to curvature quasi-topological gravities has been used to generate the regular Hayward black-hole spacetime~\cite{Bueno:2024zsx,Borissova:2026wmn,Borissova:2026krh}. This characteristic function has also been investigated in $d\geq 5$ polynomial curvature quasi-topological gravities in~\cite{Bueno:2025gjg} for FLRW cosmologies with $k=1$, which results in a qualitativaly different behavior of the scale factor as a function of time, as can be seen by considering the effective potential. Another application of the above characteristic function in $d=4$ non-polynomial gravities  can be found in~\cite{Colleaux:2019ckh}. Here we consider it as an example to illustrate some of the previous statements. \\

The equation of motion~\eqref{eq:EqMaster} can be inverted in the form
\ba\label{eq:HaywardBounce}
H^2 &=& \frac{\varrho }{1 +  \frac{ \varrho}{H_0^2} }\,,
\ea
which combined with~\eqref{eq:Rhoa} forms a complete system. \\

This choice of function $h$ results in the first scenario suggested above --- a $k=0$ de Sitter patch as the asymptotic state replacing the Big-Bang singularity, smoothly linked at $\varrho\approx H_0^2$ to a standard late-time Big-Bang model.  When $\varrho\ll H_0^2$, \eqref{eq:HaywardBounce} reduces to the standard Friedmann equation of general relativity, whereas for large values of $1/H_0^2$, the right hand side of \eqref{eq:HaywardBounce} becomes a constant. As a result, the solution for the scale factor approaches the de Sitter $k=0$ solution, 
\ba
\text{UV:} \quad \quad H^2 &\to & H_0^2\,\,\, \quad \,\,\, \Rightarrow \,\,\, \quad \,\,\, a(\tau) \,\,\to \,\, a_0 \,e^{H_0\tau}\,.
\ea
The energy density for radiation diverges as $\varrho \propto 1/a^{4}$ as $\tau \to -\infty$ and an analogue divergence occurs for any equation of state parameter $w>-1$. However, this matter
singularity is located at infinite affine distance. \\

More generally, by separating variables in~\eqref{eq:HaywardBounce},
\ba
 \sqrt{\frac{1+\frac{\varrho(a)}{H_0^2}}{a^2 \varrho(a)}}\dd{a} &=& \dd{\tau}\,,
\ea
a solution for $a$ can be found in the entire domain of $\tau$ in the form
\ba\label{eq:adeSitter}
\frac{1}{(d-1)(1+w)} \qty[2 x + \ln \qty[\abs{\frac{1 -x}{1+ x}}]]&=& H_0\qty(\tau - \tau_0)\,; \,\,\,\quad x \,\,=\,\, \sqrt{1 + \qty(\frac{a}{a_0})^{(d-1)(1+w)}}\,.\quad 
\ea
Figure~\ref{Fig:deSitter} shows the scale factor as a function of time for $d=4$ and $w = 1/3$.
Thus, a concrete solution does vindicate the previous considerations resulting from necessary but not sufficient conditions on the function $h$ and its derivative.

\subsection{Bouncing universe}\label{SecSub:Bounce}

A bounce corresponds geometrically to a strict non-zero local minimum of the function $\varphi(y) = a(\tau) r$ in the line element~\eqref{eq:Metric} and may therefore also be viewed as a wormhole throat along the time direction. The thereby implied defocusing of null geodesics ensures a violation of the null convergence condition, which is one of the central assumptions of the Penrose 1965 singularity theorem~\cite{Penrose:1964wq,Hawking:1973uf}. This is the key mechanism by which wormhole and bouncing spacetimes circumvent this and other singularity theorems~\cite{Borissova:2025msp,Borissova:2025hmj}, and as such provide viable non-singular candidate geometries~\cite{Carballo-Rubio:2024rlr,Carballo-Rubio:2019fnb,Carballo-Rubio:2019nel}.
A local minimum of $\varphi(y)$ as a function of $\tau$ occurs when $\dot{a} = 0$ and $\ddot{a} > 0$. \\

\begin{figure}[t!]
	\centering
	\includegraphics[width=0.5\textwidth]{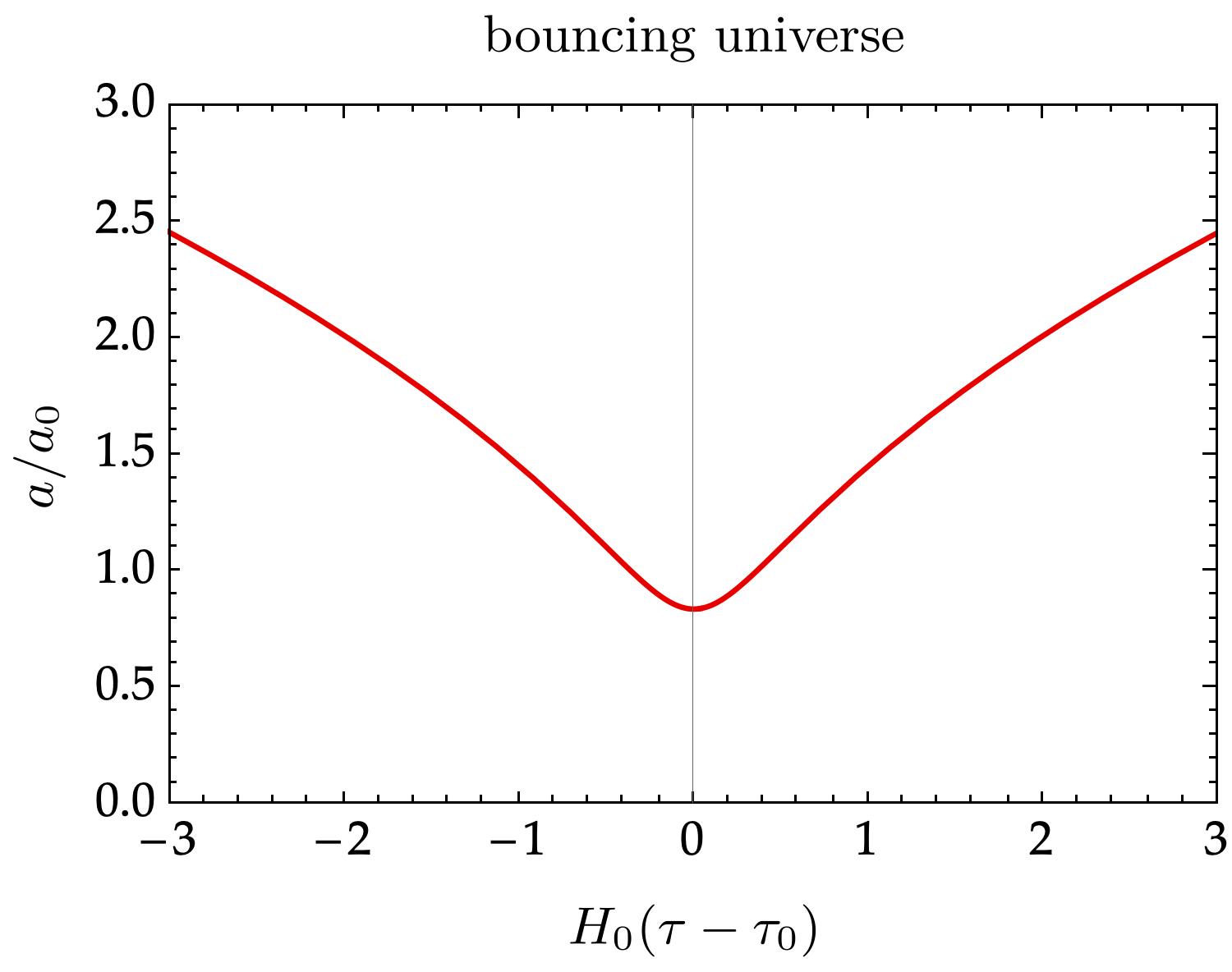}
	\caption{\label{Fig:Bounce} Scale factor as a function of time for a $d=4$ radiation-dominated FLRW universe with characteristic function $h$ in~\eqref{eq:hBounce} according to~\eqref{eq:aBounce}. The curve shows a bounce at $\tau =0$ where $\dot{a}(0)=0$ and a power-law scaling at early and late times.
	}
\end{figure}

A characteristic function resulting in a bouncing cosmology is given by combining the two branches
\ba
h\qty(H^2) &=& H_0^2 \qty[1 \pm \sqrt{1 - 2 \frac{ H^2}{H_0^2}}] \,,
\ea
see also~\cite{Ling:2025ncw}.
In this case~\eqref{eq:EqMaster} can be inverted in the form
\ba\label{eq:hBounce}
H^2 &=& \varrho \qty[1 -  \frac{\varrho }{2 H_0^2}]\,.
\ea
For $\varrho\ll H_0^2$ this equation reduces to the standard Friedmann equation of general relativity, whereas a bounce ($\dot{a}=0$) in the UV can be achieved in the limit $\varrho \to 2 H_0^2$,
\ba
\text{UV:} \quad \quad H^2 &\to & 0\,.
\ea
In general,~\eqref{eq:hBounce} can be separated in the form
\ba
\frac{\dd{a}}{\sqrt{a^2  \varrho(a) \qty[ 1 -  \frac{1 }{2 H_0^2}\varrho(a)]}} &=& \dd{\tau}\,,
\ea
and thus a solution for $a$ can be found in the form
\ba\label{eq:aBounce}
 \frac{1}{(d-1)(1+w)} \qty[4 \qty(\frac{a}{a_0})^{(d-1)(1+w)} - 2 ]^{\frac{1}{2}}&=& H_0\qty(\tau -\tau_0)\,.
\ea
For $d=4$ and $w=1/3$, a solution to the differential equation~\eqref{eq:hBounce}, imposing the initial condition $\dot{a}(\tau_0)=0$, can be stated explicitly as
\ba
a(\tau) &=& a_0 \qty[\frac{1}{2} + 4 \,H_0^2 \, \qty(\tau-\tau_0)^2 ]^{\frac{1}{4}}\,,
\ea
and is shown in Fig.~\ref{Fig:Bounce}. Again our discussion in Sec.~\ref{Sec:Acceleration} is vindicated by a specific solution. 

\subsection{Minkowski asymptotic universe}\label{SecSub:Minkowski}

Here we will provide an example for the third scenario discussed in Sec.~\ref{Sec:Acceleration} by  considering the function $h$ to be defined implicitly in terms of its inverse $h^{-1}$ satisfying $h\qty(H^2) = f\Leftrightarrow H^2 = h^{-1}\qty(f)$ as follows, 
\ba \label{eq:hMinkInv}
    H^2 &=&
    \frac{h(h-\varrho_0)^2}{\varrho_0^2}\,.
\ea
The function $h^{-1}$ for this example is actually an elementary function obtained by solving a cubic equation, but we refrain from displaying the solution here. One could also choose a power other than two for the factor $(h-\varrho_0)$ in~\eqref{eq:hMinkInv}, but the power of two is convenient for a radiation-dominated universe, and so we proceed with this choice. The important observation is that the derivative of $H^2=H^2(h)$ at $h=\varrho_0$ is zero, and conversely $h\qty(H^2)$ has infinite derivative when it hits the $H^2=0$ axis at $h(0)=\varrho_0$, cf.~the blue curve in~Fig.~\ref{Fig:PhaseDiagram}. This property distinguishes this scenario from the bouncing scenario (while in both cases $h$ must be multi-valued to accommodate the correct IR regime). \\

\begin{figure}[t!]
	\centering
	\includegraphics[width=0.5\textwidth]{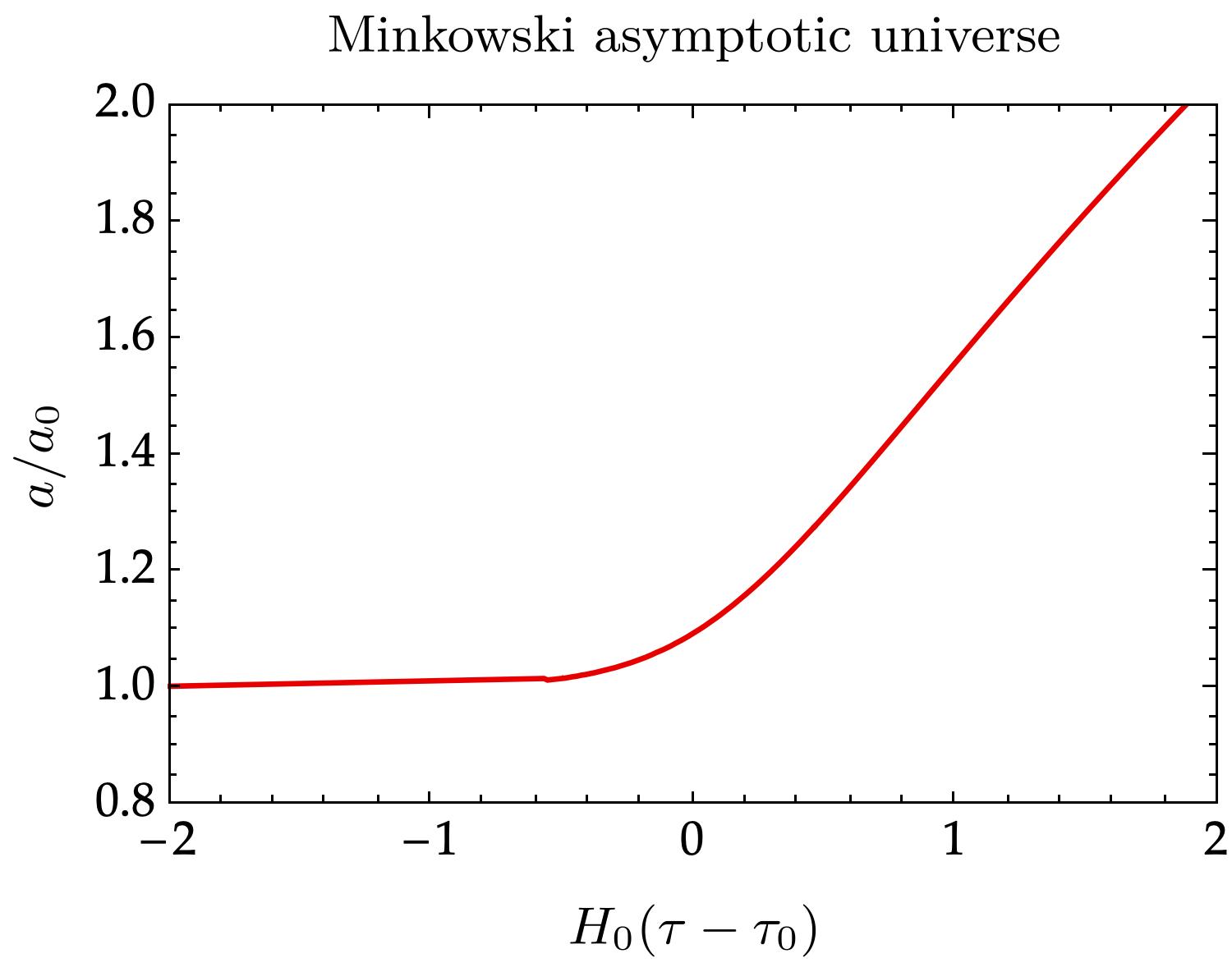}
	\caption{\label{Fig:Minkowski} Scale factor as a function of time for a $d=4$ radiation-dominated FLRW universe with characteristic function $h$ given by the inverse of~\eqref{eq:hMinkInv} according to~\eqref{eq:aMinkowski}. The curve shows a transition from a finite constant scale factor at early times to a power-law scaling at late times.}
\end{figure}

The remaining dicussion is straightforward and mimics the de Sitter example, with an important difference. For $d=4$ and $w = 1/3$, using the equation of motion~\eqref{eq:EqMaster} leads to
\ba
      H^2 &=& \frac{\varrho(\varrho-\varrho_0)^2}{\varrho_0^2}.
\ea
For $\varrho\ll \varrho_0$, we recover $H^2\approx \rho$ leading to the usual power-law scaling $a\propto \tau^{1/2}$. In the UV, as the matter energy density approaches $\varrho\rightarrow\varrho_0^-$, we have
\ba\label{H2loitering}
\text{UV:} \quad \quad     H^2 &\to & \frac{(\varrho-\varrho_0)^2}{\varrho_0}.
\ea
Taking the square root and picking the expanding branch for $\varrho<\varrho_0$,
this yields by separation of variables
\ba
 \frac{1}{4}  \log\qty[\frac{a^4-a_0^4}{c}]  &=& H_0\qty(\tau-\tau_0)\,,
\ea
where $c$ is an integration constant of dimension $\qty[a_0^4]$, and therefore
\ba
a(\tau)&=& a_0 \qty[1+ \frac{c}{a_0^4}\, e^{4H_0(\tau-\tau_0)} ]^{\frac{1}{4}}\,.
\ea
The limit $\varrho\rightarrow \varrho_0^-$ corresponds to $\tau\rightarrow -\infty$, in analogy with the de Sitter solution. We thus have in this limit
\ba
    a(\tau)& \to &a_0\qty[1+\frac{c}{4 a_0^4} e^{4 H_0 \tau}] \,\, \approx \,\, a_0\,, \quad \quad 
    \varrho \,\, \to \,\, \varrho_0^-. 
\ea
An exact solution linking the IR and UV regimes can also be found by 
separation of variables according to
\ba\label{eq:aMinkowski}
\frac{\dd{a}}{a  \sqrt{\varrho(a)} \qty[\frac{ \varrho(a)}{\varrho_0} -1]} &=& \dd{\tau}\,,
\ea
and is given by
\ba
    \frac{a^2}{2a_0^2}+\frac{1}{4}\log\qty[\frac{a^2-a_0^2}{a^2+a_0^2}] &=& H_0(\tau-\tau_0)\,.
\ea
Fig.~\ref{Fig:Minkowski} shows the scale factor as a function of $\tau$ and confirms the discussion in Sec.~\ref{Sec:Acceleration} for this particular example.

\section{Conclusions}\label{Sec:Conclusions}

In this paper we exploited the simple correspondence between the characteristic function $h(\psi)$ governing the equations of motion of the most general quasi-topological gravities constructed from the curvature tensor without covariant derivatives in $d\geq 4$ dimensions~\cite{Borissova:2026krh,Borissova:2026wmn}, and the resulting generalised Friedmann-like equation for these theories which follows from the evaluation of the two-dimensional Horndeski equations of motion on an FLRW ansatz. For $k=0$, the characteristic function
 effectively encodes the energy density as a function of the Hubble parameter --- $\rho\qty(H^2)$. 
 This allows for expressing  the generalised Friedmann equation as
\ba
H^2 &=& h^{-1}\qty(\rho)\,,
\ea
assuming $h$ has an inverse. This relation
provides a straightforward way to identify candidates realising the three cosmological singularity-resolving scenarios discussed in this work.\\

Aside from issues related to the amount of proper time spent in different regions of the curves in Fig.~\ref{Fig:PhaseDiagram}, the following constitute necessary conditions for these different  scenarios.

\begin{itemize}

\item \textbf{de Sitter asymptotic universe:} 
The function $h\qty(H^2)$ must develop a singularity at $H^2 = H_0^2$, corresponding to the square of the UV Hubble parameter characterising the de Sitter regime. For a normal fluid (e.g., radiation), the energy density $\rho$ must diverge as $\tau \to -\infty$, implying that $h\qty(H^2)$ diverges at this point.

\item \textbf{Bouncing universe:} 
The function $h\qty(H^2)$ must intersect the $h$-axis defined by $H^2=0$ twice --- once in the IR and once in the UV at the bounce, where $H^2=0$ but $\rho=\rho_{\rm max}$. Given the IR behaviour required to recover general relativity at late times, i.e., $h\qty(H^2)\sim H^2$, this requires $h\qty(H^2)$ to be double-valued. One must then verify that the time evolution indeed carries the universe from one branch to the other, and subsequently back into a contracting pre-bounce phase.

\item \textbf{Minkowski asymptotic universe:} 
As in the bouncing scenario, $h\qty(H^2)$ must approach the $h$-axis in the UV so that $H^2$ aymptotically vanishes as $\rho$ reaches a maximum value $\rho_{\rm max}$. In addition, since a static universe requires $\ddot a \rightarrow 0$, Eq.~\eqref{ddotamain} implies that $h'(0)$ must diverge at this point. This is a necessary (albeit not sufficient) condition for the universe to asymptote to Minkowski spacetime at early times. Unlike the de Sitter scenario, the energy density remains finite and the curvature vanishes, thereby avoiding trans-Planckian issues.

\end{itemize}

These schematic considerations must then be complemented by explicit calculations to verify that the full dynamics indeed realise the anticipated behaviour. In this paper we have carried out this analysis for specific exemplary choices of $h\qty(H^2)$. In the $h\qty(H^2)$ diagram these three cases appear as closely related variants of one another. In all scenarios $h'$ must diverge at some finite value of $H^2$, effectively capping the curvature, but the detailed behaviour distinguishes the models. \\

In the de Sitter case the function $h$ itself diverges, whereas in the other two cases it continues smoothly onto another branch of a multi-valued function. 

The bouncing and Minkowski scenarios both re-intersect the $h$ axis at $h(0)>0$, corresponding to $H^2=0$ at finite $\rho_{\rm max}$, but they differ in the behaviour of the derivative: in the bouncing case $h'(0)$ remains finite, whereas in the Minkowski case $h'(0)$  diverges. This distinction is crucial, as it suggests that the asymptotic Minkowski universe may be interpreted as a ``failed bounce'' --- a bounce pushed to infinite affine distance and therefore lacking a preceding contracting phase. It would be interesting to investigate whether such a limit is realised in loop quantum cosmology.

An early Minkowski origin has several advantages over a de Sitter start. Unlike the de Sitter case, the matter singularity is removed even at infinite affine distance. It also avoids the trans-Planckian problem of the regular de Sitter model, since $\rho$ cannot grow arbitrarily large. In this sense, the scenario exhibits a form of asymptotic freedom (rather than generic asymptotic safety) --- in the UV gravity effectively switches off, yielding a regular near-Minkowski universe even when $\rho$ is large. The curvature remains zero, while matter scalars can become large but never diverge. Besides being the most original scenario explored in this paper, it also appears to possess the most desirable physical properties.

Finally, let us emphasise that the considerations here apply a priori only to the particular case of an FLRW universe with spatial curvature $k=0$. For other values of $k$, bouncing and eternal cosmologies such as the ones considered here may not require a multi-valued characteristic function.

\begin{acknowledgments}

We thank Ra\'ul Carballo-Rubio  for  related discussions and Aimeric Coll\'eaux for written communication. Both authors acknowledge support by STFC Consolidated Grant ST/T000791/1.

\end{acknowledgments}

\bibliographystyle{jhep}
\bibliography{references}

\end{document}